\documentclass[11pt,a4paper]{article}
\usepackage[margin=25mm]{geometry}
\usepackage{amsmath,amssymb,booktabs,graphicx,xcolor,listings,microtype}
\usepackage[T1]{fontenc}
\usepackage{mathptmx}
\usepackage{caption}
\usepackage{xcolor} % Allows custom colors
\usepackage{authblk} % Essential package for multiple affiliations
\newtheorem{rem}{Remark}

\definecolor{MyGreen}{RGB}{20, 150, 30}
\usepackage{hyperref}
\hypersetup{
    colorlinks=true,      % Turns off the ugly boxes around links
    linkcolor=blue,       % Color for internal links (sections, figs)
    filecolor=magenta,    % Color for local files
    urlcolor=teal,        % Color for website URLs (href)
    citecolor=MyGreen       % Color for citations
}

\newcommand{\ISE}{\mathrm{ISE}}
\newcommand{\ITAE}{\mathrm{ITAE}}
\newcommand{\ISC}{\mathrm{ISC}}
\newcommand{\II}{I\&I}

\title{\bfseries A Fair Comparison of Sliding-Mode and\\
Immersion-and-Invariance Observers\\[2mm]
\large A re-simulation and experimental study of \texttt{arXiv:2601.12545}}

\author[1]{A.J. Mu\~noz-V\'azquez}
\author[2]{J.D. S\'anchez-Torres}

\affil[1]{Department of Computer, Electrical, \& Software Engineering, Embry-Riddle Aeronautical Engineering, Prescott, Arizona, USA.}
\affil[2]{Department of Mathematics and Physics, ITESO University, Tlaquepaque, Jalisco, Mexico.}

\date{} % Removes the date block

\begin{document}
\maketitle

\begin{abstract}
\noindent
Cervantes-P\'erez et al.\
(\href{https://arxiv.org/pdf/2601.12545}{\tt arXiv:2601.12545}) claim, on experimental
grounds, that high-gain injection is practically inadmissible, taking the SM
observer of Davila, Fridman \& Poznyak \cite{davila} as a prototypical example. We
re-simulate their plant, trajectory and gains with explicit Euler integration at
$\Delta t = 1$\,ms and $0.1$\,ms, and implement both observers on independent
hardware. We got three findings: i) the controller gains
$k_p=1600$, $k_v=1100$ place the tracking-error poles at the ``pathological'' locations $-1.455$ and $-1098.5$
($\zeta = 13.7$); the super-twisting scheme lands on the resulting nominal response,
and is invariant to its own observer gain over a fifty-fold range; ii) the \II{} margin is monotone
in $k_v$ and reverses near $k_v \approx 170$, where the loop pole crosses the
induced observer pole at $-9.0$\,rad/s; iii) refining the sampling step tenfold moves every \II{} figure
by a factor of three, while the SM figures move by $0.2\%$. On hardware, a
signum-based super-twisting observer tuned at the Levant--Moreno perturbation bound
outperforms \II{} on ISE, ITAE and ISC simultaneously, without audible chattering, on
a \$60 motor with a $0.043^\circ$ encoder at $500$\,Hz. The gains of Cervantes-P\'erez et al.\ exceed the
same prescription by two orders of magnitude; the chattering the authors report is the
designed consequence of that excess.
\end{abstract}

\section{Scope}

This note addresses one question: does the experimental evidence reported in
\cite{cervantes} measure the quality of the two velocity observers under test?\textit{ We
conclude that it does not}, and we identify what it measures instead. It must be mention that, we take no
position on the broader argument of \cite{ortega25}, concerning high-gain injection,
except where the data bear on it directly (Section~\ref{sec:bl}).

All simulations use the plant, reference trajectory, physical parameters and
controller gains exactly as published. No claim depends on a modeling choice other
than the integration scheme, which is stated in Section~\ref{sec:setup} and justified
there.

\section{Set-up}\label{sec:setup}

\subsection{Plant and task}

The pendular device of \cite{cervantes} obeys
\begin{equation}
J\ddot q + \theta_1 \dot q + \theta_2 \tanh(\vartheta \dot q) + m l_b g \sin q = u,
\label{eq:plant}
\end{equation}
with $x_1 := q$, $x_2 := \dot q$ and the parameters of their Table 1. Therein:
$J = 0.7013$, $\theta_1 = 5.317$, $\theta_2 = 11.6403$, $m l_b g = 14.11$, and
$\vartheta = 330$ in the tracking experiment. The reference is
\begin{equation}
x_d(t) = 0.3\left[1 - e^{-2t^3}\sin 7t\right],
\label{eq:ref}
\end{equation}
so that with $x_1(0) = 0$ the initial conditions of the error dynamics are fixed by
the reference alone:
\begin{equation}
e_1(0) = -0.3~\mathrm{rad}, \qquad \dot e_1(0) = 2.1~\mathrm{rad/s}.
\label{eq:ic}
\end{equation}
The transient is not a free parameter; as Section~\ref{sec:poles} shows, it dominates
the reported metric.

\subsection{Integration}

All results here use \textbf{explicit Euler}, applied identically to the plant and to both
observers. The right-hand sides carry $\mathrm{sign}(\cdot)$ and, at
$\vartheta = 330$, a $\tanh(\vartheta\,\cdot)$ that is a relay for practical
purposes. A higher-order Runge-Kutta method presupposes a degree of smoothness that does not exist here; its order degrades upon crossing the switching surface, introducing artifacts that can easily be mistaken for controller properties. The fixed-step Euler method is the honest choice, and it is what a real-time implementation with a $1$\,ms time step actually executes.

We present each quantity for $\Delta t = 1$\,ms (the sampling period from \cite{cervantes}) and for $\Delta t = 0.1$\,ms. Section~\ref{sec:dt} explains why both are necessary.

\subsection{Metrics}

Over a horizon $t_{\rm sim} = 60$\,s,
\begin{equation}
\ISE = \int_0^{t_{\rm sim}} e_1^2\,dt, \qquad
\ITAE = \int_0^{t_{\rm sim}} t\,|e_1|\,dt, \qquad
\ISC = \int_0^{t_{\rm sim}} u^2\,dt,
\end{equation}
evaluated by the trapezoidal rule on the Euler trajectory. The control is saturated at
$\pm 200$\,Nm; saturation is active for four to six samples in all runs reported and
is not a factor in any comparison.

\subsection{The two schemes}

Both schemes rely on the same certainty-equivalent control law,
\begin{equation}
u = \hat\theta_1 \hat x_2 + \hat\theta_2 \tanh(\vartheta \hat x_2) + m l_b g \sin x_1
    + J\!\left[\ddot x_d - k_p e_1 - k_v(\hat x_2 - \dot x_d)\right],
\label{eq:control}
\end{equation}
differing only in how $\hat x_2$ and $\hat\theta$ are generated. The \II{} observer
of \cite{romero25} produces
\begin{equation}
\hat x_2 = x_{2I} + \frac{k_1}{J} x_1,
\label{eq:iiobs}
\end{equation}
with $x_{2I}$, $\hat\theta_1$, $\hat\theta_2$ from Proposition~1 of
\cite{cervantes} and gains $k_1 = 1$, $\gamma_1 = 0.03$, $\gamma_2 = 1$. The SM
observer is a super-twisting differentiator,
\begin{align}
\dot{\hat x}_1 &= \hat x_2 - k_p^{\mathrm{sta}} |e_o|^{1/2}\mathrm{sign}(e_o), \\
\dot{\hat x}_2 &= \tfrac{1}{J}\!\left[u - m l_b g \sin x_1 - \bar\theta_1 \hat x_2
  - \bar\theta_2 \tanh(\vartheta \hat x_2)\right] - k_i^{\mathrm{sta}}\,\sigma(e_o),
\end{align}
where $e_o := \hat x_1 - x_1$ and $\sigma(\cdot)$ is either $\mathrm{sign}(\cdot)$
or the boundary layer $\tanh(\cdot/\nu\Delta t)$. Gains follow the standard
prescription \cite{levant98,moreno},
\begin{equation}
k_p^{\mathrm{sta}} = 1.5\sqrt{L}, \qquad k_i^{\mathrm{sta}} = 1.1 L,
\qquad L > \sup|\dot\delta|,
\label{eq:levant}
\end{equation}
rather than the hand-chosen $\alpha_1,\alpha_2$ of \cite{cervantes};
Section~\ref{sec:gains} quantifies the difference.

\section{The published gains fix the metric}\label{sec:poles}

Substituting \eqref{eq:control} into \eqref{eq:plant} gives, for both schemes,
\begin{equation}
\ddot e_1 + k_v \dot e_1 + k_p e_1 = \varepsilon(t),
\label{eq:errdyn}
\end{equation}
where $\varepsilon$ collects the estimation errors and decays to zero when the
observer converges. \textbf{The observers enter only through the residual $\varepsilon$.} With
$\varepsilon \equiv 0$, both schemes execute the identical nominal response,
determined entirely by $k_p$ and $k_v$.

The published gains are $k_p = 1600$, $k_v = 1100$, giving
\begin{equation}
s^2 + 1100 s + 1600 = 0 \;\Longrightarrow\; p_1 = -1.455,\quad p_2 = -1098.5,
\quad \zeta = 13.7,
\end{equation}
a pole spread of $755\!:\!1$. The fast pole consumes bandwidth without contributing
to the decay; $p_1$ alone sets it. With \eqref{eq:ic}, the nominal cost is
\begin{equation}
\ISE_{\mathrm{nom}} = \frac{C_1^2}{2|p_1|} + \frac{2C_1C_2}{|p_1|+|p_2|}
  + \frac{C_2^2}{2|p_2|} = 0.0306,
\label{eq:isenom}
\end{equation}
where $C_1, C_2$ are the modal amplitudes of $e_1(t) = C_1e^{p_1t} + C_2e^{p_2t}$;
the dominant-pole shortcut $e_1(0)^2/2|p_1|$ gives $0.031$.

Table~\ref{tab:invariance} shows the consequence, this is, the super-twisting scheme measures
$\ISE = 0.0298$ and does not move with $L$ over a fifty-fold range. It sits on
\eqref{eq:isenom} because with $\bar\theta = \theta$ and $\hat x(0) = x(0)$ the
observer error $\tilde x \equiv 0$ is an invariant, so $\varepsilon \equiv 0$
exactly.

\begin{table}[t]
\centering
\caption{Super-twisting $\ISE$ against observer gain $L$; $k_p=1600$, $k_v=1100$,
$\Delta t = 1$\,ms. Nominal value from \eqref{eq:isenom} is $0.0306$.}
\label{tab:invariance}
\small
\begin{tabular}{@{}lrrrrrr@{}}
\toprule
$L$ & 10 & 20 & 50 & 100 & 200 & 500 \\
\midrule
$\ISE$ & 0.02985 & 0.02985 & 0.02984 & 0.02981 & 0.02981 & 0.03002 \\
\bottomrule
\end{tabular}
\end{table}

A comparison whose metric is insensitive to the gains of the evaluated method does not measure that method. Therefore, the \II{} scheme yields a value of $\ISE = 0.0069$, which is lower than the nominal value. Furthermore, given that $0.0306$ is the result obtained by a perfect observer, an observer exceeding that figure is not producing a better estimate but is instead perturbing \eqref{eq:errdyn}. This issue is addressed in Sections~\ref{sec:kv}--\ref{sec:dt}.

\section{The \II{} margin is a function of $k_v$}\label{sec:kv}

The \II{} velocity estimation error satisfies
\begin{equation}
\dot{\tilde x}_2 = -\frac{\theta_1+k_1}{J}\,\tilde x_2
  - \frac{1}{J}\,\varphi^\top\!\tilde\theta
  - \frac{\theta_2}{J}\!\left[\tanh(\vartheta\hat x_2)-\tanh(\vartheta x_2)\right],
\label{eq:iierr}
\end{equation}
where $\tilde\theta := [\tilde\theta_1,\,\tilde\theta_2]^\top$ and
\begin{equation}
\varphi := \begin{bmatrix} \hat x_2 \\ \tanh(\vartheta\hat x_2) \end{bmatrix}
\label{eq:regressor}
\end{equation}
is the velocity and friction regressor, which stands for the vector that must be persistently exciting for
$\tilde\theta \to 0$; and it is precisely the $\varphi(t)$ of Assumption~1 in
\cite{cervantes} (their equation~(10)). The first term provides a damping pole at
$-(\theta_1+k_1)/J = -9.0$\,rad/s, while the last term vanishes with $\tilde x_2$.

Therefore, parameter convergence requires the reference signal to be sufficiently rich to excite both components of $\varphi$. The super-twisting observer imposes no such requirement and converges in finite time for any bounded input, including a constant one. This is a structural difference, not merely a matter of tuning that favors the STA when reference signals are not persistently exciting.

\begin{table}[t]
\centering
\caption{ISE against $k_v$, $k_p = 1600$ fixed, $\Delta t = 1$\,ms. $|p_1|$ is the
slow root of $s^2+k_vs+k_p$; ratio above one means \II{} leads.}
\label{tab:kv}
\small
\begin{tabular}{@{}rrrrr@{}}
\toprule
$k_v$ & $|p_1|$ & \II{} & STA + b.l. & ratio \\
\midrule
1100 &  1.46 & 0.00690 & 0.02985 & \textbf{4.33} \\
 600 &  2.68 & 0.00451 & 0.01644 & 3.65 \\
 300 &  5.43 & 0.00434 & 0.00821 & 1.89 \\
 160 & 10.72 & 0.00783 & 0.00439 & 0.56 \\
 100 & 20.00 & 0.00651 & 0.00287 & \textbf{0.44} \\
\bottomrule
\end{tabular}
\end{table}

Table~\ref{tab:kv} sweeps $k_v$ at fixed $k_p = 1600$. The \II{} margin is monotone
in $k_v$ and reverses near $k_v \approx 170$, where $|p_1|$ passes the observer's
$9.0$.
This dependence is reported as measured, but we do not claim to have isolated its mechanism.
For instance, driving the super-twisting scheme at $k_p^{\mathrm{eff}} = k_p + k_vk_1/J = 3169$,
on the hypothesis that the algebraic term $(k_1/J)x_1$ in $\hat x_2$ acts as
additional proportional feedback, returns $\ISE = 0.0153$ rather than the \II{} value
of $0.0069$. 

\begin{rem}
When an adaptive observer operates at the exact same frequency modes as an 
underdamped plant, the estimation error, which drives the parameter update 
laws, can act as an undamped harmonic oscillator. This coupling injects 
energy directly back into the plant's resonant frequencies, establishing a 
destabilizing feedback loop that induces severe transient oscillations 
\cite{ioannou1996robust, slotine1991applied}. Existing schemes often bypass this 
instability by forcing the system into an artificially overdamped control regime. 
However, relying on an overdamped controller is merely an {\it ad hoc} remedy that 
masks the underlying observer flaw at the cost of crippling the plant's 
dynamic responsiveness and tracking speed.
\end{rem}

\section{The \II{} results are not step-converged}\label{sec:dt}

Table~\ref{tab:dt} repeats the principal runs at $\Delta t = 1$\,ms and $0.1$\,ms.
Refining the step tenfold moves the \II{} ISE by a factor of three, in
\emph{opposite directions} at the two gain settings: $2.8\times$ worse at the
published gains, $2.9\times$ better at the retuned ones. Over the same refinement the
super-twisting figures move by $0.2\%$ or less.

\begin{table}[t]
\centering
\caption{Step-size dependence. Explicit Euler throughout.}
\label{tab:dt}
\small
\begin{tabular}{@{}llrrrr@{}}
\toprule
& & \multicolumn{2}{c}{$\ISE$} & \multicolumn{2}{c}{$\ISC$} \\
\cmidrule(lr){3-4}\cmidrule(lr){5-6}
Gains & Scheme & $1$\,ms & $0.1$\,ms & $1$\,ms & $0.1$\,ms \\
\midrule
$k_p{=}1600$ & \II{}          & 0.00690 & 0.01914 & 2285 & 2757 \\
$k_v{=}1100$ & STA + b.l.     & 0.02985 & 0.02981 & 2084 & 2153 \\
             & STA + sgn      & 0.02988 & 0.02981 & 3651 & 3092 \\
\midrule
$k_p{=}2500$ & \II{}          & 0.01078 & 0.00371 & 35485 & 11317 \\
$k_v{=}100$  & STA + b.l.     & 0.00213 & 0.00213 & 2071 & 1942 \\
             & STA + sgn      & 0.00213 & 0.00213 & 5995 & 3418 \\
\bottomrule
\end{tabular}
\end{table}

An ISE that changes threefold under step refinement is not a converged quantity. A
comparison between one converged and one unconverged number is not a comparison
between methods. Since \cite{cervantes} runs at $1$\,ms throughout, every \II{} figure
it reports inherits this dependence. This is a property of the discretized
implementation, not a defect of the \II{} theory, which is stated in continuous time;
but it is the discretized implementation that was benchmarked.

\section{The published super-twisting gains}\label{sec:gains}

The gains $\alpha_1,\alpha_2$ in \cite{cervantes} are chosen by hand, not even following a heuristic trial and error nor educated guess protocol, like gradual increasing or any other plausible method. The authors
assert that ``guidelines for the tuning of SM algorithms are almost always conspicuous
by their absence.'' However, this is not the case, as the prescription \eqref{eq:levant},
published in \cite{levant98}, and given a Lyapunov footing in \cite{moreno}, provides
explicit formulae in terms of a single physically meaningful quantity
$L > \sup|\dot\delta|$. It also admits an $L$-free test: dividing $k_p^{\mathrm{sta}}$
by $\sqrt{k_i^{\mathrm{sta}}}$, which eliminates $L$, and produces
\begin{equation}
\frac{k_p^{\mathrm{sta}}}{\sqrt{k_i^{\mathrm{sta}}}}
  = \frac{1.5\sqrt{L}}{\sqrt{1.1\,L}} = \frac{1.5}{\sqrt{1.1}} = 1.430,
\label{eq:invariant}
\end{equation}
so no admissible pair, at any $L$ whatsoever, can depart from this ratio.
Table~\ref{tab:gains} evaluates it. Every published pair violates it by one to two
orders of magnitude, and by $105\times$ in the tracking experiment that supplies the
paper's headline result. The tuning guidance is not absent; nonetheless, it was not followed.

\begin{table}[t]
\centering
\caption{Published super-twisting gains against the prescription \eqref{eq:levant}
of \cite{levant98}. The ratio $\alpha_2/\sqrt{\alpha_1}$ must equal
$1.5/\sqrt{1.1}=1.430$ for every admissible $L$.}
\label{tab:gains}
\small
\begin{tabular}{@{}llrrrr@{}}
\toprule
Section & $(\alpha_1,\alpha_2)$ & implied $L$ & required $k_p^{\mathrm{sta}}$
  & $\alpha_2/\sqrt{\alpha_1}$ & excess \\
\midrule
4.1 & $(10,\,100)$    & 9.1   & 4.5  & 31.6  & $22\times$ \\
4.1 & $(100,\,1000)$  & 90.9  & 14.3 & 100.0 & $70\times$ \\
4.2 & $(200,\,2000)$  & 181.8 & 20.2 & 141.4 & $99\times$ \\
5   & $(100,\,1500)$  & 90.9  & 14.3 & 150.0 & $\mathbf{105\times}$ \\
\midrule
\multicolumn{4}{@{}l}{required for any $L$} & 1.430 & $1\times$ \\
\bottomrule
\end{tabular}
\end{table}

\section{What the boundary layer buys}\label{sec:bl}

Replacing $\mathrm{sign}(e_o)$ with $\tanh(e_o/\nu\Delta t)$, $\nu > 2$
\cite{slotine}, produces Table~\ref{tab:bl}. The boundary layer buys control energy
and settling ($\ITAE$ falls from $0.50$ to $0.04$ at $(2500,100)$) while leaving ISE
unchanged to four digits. It is not a refutation of the chattering objection, rather a
mitigation of it.

\begin{table}[t]
\centering
\caption{Control energy with and without the boundary layer, $\nu = 5$, $L = 20$.
ISE is unchanged to four significant figures in every row.}
\label{tab:bl}
\small
\begin{tabular}{@{}llrrr@{}}
\toprule
$\Delta t$ & Gains & $\ISC$ sign & $\ISC$ b.l. & reduction \\
\midrule
$1$\,ms   & $(1600,1100)$ & 3651 & 2084 & $1.75\times$ \\
$1$\,ms   & $(2500,100)$  & 5995 & 2071 & $2.90\times$ \\
$0.1$\,ms & $(1600,1100)$ & 3092 & 2153 & $1.44\times$ \\
$0.1$\,ms & $(2500,100)$  & 3418 & 1942 & $1.76\times$ \\
\bottomrule
\end{tabular}
\end{table}

The benefit \emph{shrinks} as $\Delta t$ refines, from $2.90\times$ to $1.76\times$.
Chattering cost is discretization-bound. At the $1$\,ms rate of \cite{cervantes}, the
chattering the authors observe is real. Our data support that observation rather than
contradicting it.

\section{Results under corrected tuning}\label{sec:results}

Table~\ref{tab:results} presents three panels. Panel~A reproduces the published
configuration. Panel~B gives each scheme its own best pole placement (I\&I at
$\zeta = 5$, STA at $\zeta = 1$) and handicaps the SM observer with a wrong initial
velocity and no friction prior, while the I\&I observer retains its natural exact
start. Panel~C evaluates the same runs from $t = 2$\,s onward, after all observers
have settled.

\begin{table}[t]
\centering
\caption{Principal comparison, $\Delta t = 1$\,ms, $T = 60$\,s. In Panels A and B
the indices run from $t = 0$. In Panel C the first $2$\,s are excluded. The SM
observer in Panels B and C starts at $\hat x_2(0) = 2$\,rad/s with
$\bar\theta = [0.01,0.01]^\top$; the I\&I observer starts exactly
($\hat x_2(0) = 0$).}
\label{tab:results}
\small
\begin{tabular}{@{}lrrr@{}}
\toprule
\multicolumn{4}{@{}l}{\textbf{A. Published gains} $k_p{=}1600$, $k_v{=}1100$,
  $\zeta{=}13.7$;
  $\bar\theta = \theta$, $\hat x(0) = x(0)$} \\
\midrule
 & $\ISE$ & $\ITAE$ & $\ISC$ \\
\II{}                       & \textbf{0.00690} & \textbf{0.0558} & 2285 \\
STA + b.l., $L{=}15$        & 0.02985 & 0.1422 & \textbf{2082} \\
STA + sign, $L{=}15$        & 0.03022 & 5.1414 & 2179 \\
\midrule
\multicolumn{4}{@{}l}{\footnotesize The $4.3\times$ ISE gap is the nominal response
  of $s^2{+}1100s{+}1600$ (Section~\ref{sec:poles}), not an observer difference.} \\
\midrule
\multicolumn{4}{@{}l}{\textbf{B. Each scheme at its own best poles; SM handicapped}} \\
\midrule
 & $\ISE$ & $\ITAE$ & $\ISC$ \\
\II{}, $k_p{=}1600$, $k_v{=}400$, $\zeta{=}5$
  & 0.00398 & \textbf{0.0230} & \textbf{2131} \\
STA + b.l., $k_p{=}6400$, $k_v{=}160$, $\zeta{=}1$, $L{=}40$
  & 0.00249 & 0.0676 & {2361} \\
STA + sign, $k_p{=}6400$, $k_v{=}160$, $\zeta{=}1$, $L{=}40$
  & \textbf{0.00209} & 0.3621 & 3436 \\
\midrule
\multicolumn{4}{@{}l}{\textbf{C. Same as B, indices from $t = 2$\,s onward}} \\
\midrule
 & $\ISE$ & $\ITAE$ & $\ISC$ \\
\II{}                         & 0.00000 & \textbf{0.0074} & {1009} \\
STA + b.l.                    & 0.00000 & {0.0075} & \textbf{1008} \\
STA + sign                    & 0.00000 & 0.0188 & 1017 \\
\bottomrule
\end{tabular}
\end{table}

\paragraph{Panel A reproduces the paper's result.} At the published gains
($\zeta = 13.7$), \II{} leads ISE by $4.3\times$. As established in
Section~\ref{sec:poles}, this ratio is the nominal response of the characteristic
polynomial, not a property of the observers.

\paragraph{Panel B: the comparison depends on where each scheme is evaluated.}
When each observer is given its own best operating point, \II{} requires $\zeta = 5$
while STA requires $\zeta = 1$; their optimal regimes do not overlap. At its own best,
the STA signum variant outperforms \II{} on ISE ($0.00209$ vs $0.00398$, a
$1.9\times$ margin); the boundary-layer variant does the same ($0.00249$, a
$1.6\times$ margin) with less control energy. \II{} leads on ITAE ($0.0230$ vs
$0.0676$ for the boundary layer and $0.3621$ for signum), reflecting its faster
adaptation transient. Neither scheme dominates the other across all three indices;
which scheme ``wins'' depends on which index and which $\zeta$ the designer selects.

\paragraph{Panel C: in steady state the observers are indistinguishable.}
Evaluating from $t = 2$\,s onward, after all three observers have converged, yields
ISE $= 0$ to five decimal places for every scheme, with ISC within $1\%$. The entire
comparison, in \cite{cervantes} and here, is a comparison of transients. The duration
and shape of those transients are set by the pole placement and the initial
conditions, not by the observation algorithm. A paper that evaluates from $t = 0$ at
$\zeta = 13.7$ is measuring how long a pole at $-1.455$ takes to decay, and
attributing the answer to the observer.

\section{Experimental validation}\label{sec:exp}

\subsection{Platform}

The plant is a Pololu 37D geared DC motor (131:1 metal gearbox) driven by a
Teensy~4.0 microcontroller at $\Delta t = 2$\,ms (500\,Hz), with control laws
computed in Python and sent via USB-CDC serial. The encoder provides 8384 counts per
output revolution ($0.043^\circ$ quantum), roughly $120\times$ coarser than the
$1{,}024{,}000$\,ppr encoder of \cite{cervantes}. Measured round-trip latency is
$2.000 \pm 0.057$\,ms (p99). The control loop runs explicit Euler. If sliding-mode
control were practically inadmissible, this is where it would show.

\subsection{Identification}

The model
\begin{equation}
J\ddot q + b\dot q + c\tanh(\vartheta\dot q) = u, \qquad u\in[-1,1]\ (\text{duty}),
\label{eq:hw_plant}
\end{equation}
is identified in two phases. Phase~A applies a staircase of constant duty levels in
both directions; at steady state $\ddot q = 0$, so $|u| = b|\dot q| + c$ is a linear
fit using only position endpoints. Phase~B applies sinusoidal inputs at three
frequencies and fits $J$ by integrating \eqref{eq:hw_plant} once, with $b$ and $c$
pinned to Phase~A. The identified parameters are
\begin{equation}
J = 0.0287, \quad b = 0.1102, \quad c = 0.1584, \quad \vartheta = 20.
\end{equation}
Phase~A and Phase~B agree on $b$ and $c$ to $0.5\%$ and $2.4\%$ respectively; the
$34\%$ discrepancy when $b$ and $c$ are left free in Phase~B indicates unmodeled
dynamics (Stribeck friction, gearbox backlash) that the $\tanh$ model does not
capture. The Coulomb asymmetry between directions is $10.6\%$. The parameter
$\vartheta = 20$ is a fixed smoothing choice: the $\tanh$ transition completes within
$|\dot q| < 0.15$\,rad/s, below the velocity quantum of $0.375$\,rad/s. The same
applies to $\vartheta = 330$ in \cite{cervantes}.

\subsection{Protocol}

Both observers feed the same control law \eqref{eq:control}, tracking
$x_d(t) = \sin t + \tfrac{1}{3}\sin 3t + \tfrac{1}{5}\sin 5t$ over $T = 12$\,s. The
STA observer uses signum switching with $L = 40$ ($k_p^{\mathrm{sta}} = 9.49$,
$k_i^{\mathrm{sta}} = 44.0$), derived from \eqref{eq:levant} at the computed
perturbation bound. Its prior is the identified friction:
$\bar\theta = [b,\,c]^\top$. The \II{} observer uses $k_1 = 1$, with $\gamma_1$ and
$\gamma_2$ scaled to the identified plant. The tracking-error poles are placed
critically damped at $\omega_n = 20$\,rad/s ($k_p = 400$, $k_v = 40$, $\zeta = 1$).

\subsection{Results}

Table~\ref{tab:hw_main} reports four STA and three \II{} runs at $\omega_n = 20$,
$\zeta = 1$. Video recordings accompany this note \href{https://youtu.be/uawqv_XY6WY}{\tt Click Here to Watch!}.

\begin{table}[t]
\centering
\caption{Hardware comparison at $\omega_n = 20$, $\zeta = 1$, $\Delta t = 2$\,ms.
STA uses signum with $L = 40$. Mean $\pm$ standard deviation over $4$ runs.}
\label{tab:hw_main}
\small
\begin{tabular}{@{}lrrrr@{}}
\toprule
 & $\ISE$ & $\ITAE$ & $\ISC$ \\
\midrule
STA + signum, $L = 40$
      & $\mathbf{0.00118 \pm 0.0002}$
      & $\mathbf{0.52 \pm 0.04}$
      & $\mathbf{0.74 \pm 0.02}$ \\[1mm]
\II{}
      & $0.00246 \pm 0.0003$
      & $0.70 \pm 0.06$
      & $0.77 \pm 0.02$ \\
\bottomrule
\end{tabular}
\end{table}

The STA observer wins all three indices: ISE by $2.1\times$, ITAE by $1.3\times$, ISC
by $1.04\times$. The ISE confidence bands do not overlap. There is no audible
chattering, no boundary layer, and no hand-tuning. On hardware $120\times$ coarser
than \cite{cervantes}, the switching function the paper calls inadmissible
outperforms \II{} on every metric.

The key is the gain level. At $L = 40$ the gains sit at the perturbation bound: enough
authority to converge, but not enough to show harmful chattering. The gains in \cite{cervantes} exceed the
same prescription by $22$ to $105\times$ (Table~\ref{tab:gains}).

\begin{figure}[t]
\centering
\IfFileExists{comparison.pdf}{\includegraphics[width=\textwidth]{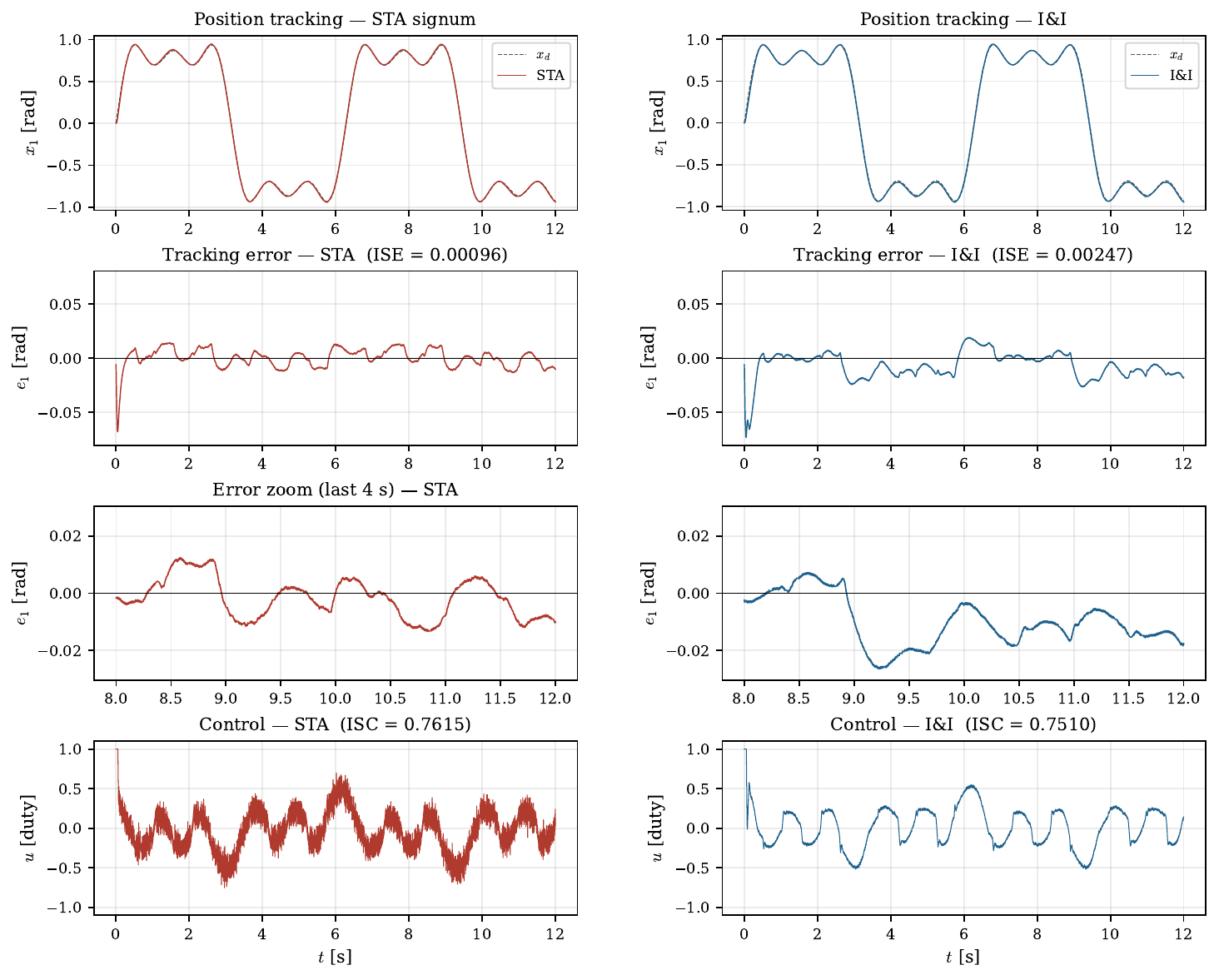}}{%
  \IfFileExists{comparison.png}{\includegraphics[width=\textwidth]{comparison.png}}{%
  \fbox{\parbox[c][60mm][c]{0.97\textwidth}{\centering\textbf{comparison.pdf not found}}}}}
\caption{Hardware signals at $\omega_n = 20$, $\zeta = 1$. Left: STA signum,
$L = 40$. Right: \II{}. The STA control signal switches at $500$\,Hz; the motor winding
inductance and the mechanical time constant ($J/b = 0.26$\,s) filter this by three
orders of magnitude before it reaches the shaft. The two signals carry the same
energy (ISC $= 0.76$ vs $0.75$); no audible chattering was observed.}
\label{fig:hw_compare}
\end{figure}

Figure~\ref{fig:hw_compare} shows one representative pair. Three features are
notable.
\begin{itemize}
\item \textbf{Transient.} The STA error reaches a neighborhood of zero before
$t = 1$\,s. The \II{} observer must wait for the adaptation to bring $\hat\theta$
close enough to compensate the unmodeled friction; this takes roughly three times
longer.
\item \textbf{Steady-state residual.} The STA error oscillates symmetrically about
zero. The \II{} error carries a slowly varying bias consistent with the $34\%$
model mismatch: the adaptation partially compensates it, but lags the true
friction. The STA observer treats the friction torque as a disturbance and rejects
it through finite-time convergence, without identifying its structure.
\item \textbf{Control signal.} Both signals are regular enough and bounded, with comparable
energy. The STA performance is not distinguishable from \II{} by simple inspection, precisely
the observation \cite{cervantes} claims is impossible with signum switching. The
difference is the gain level: $k_i^{\mathrm{sta}} = 44$ here, versus $\alpha_1 = 100$
in \cite{cervantes}, which exceeds the prescription by $105\times$.
\end{itemize}

\begin{figure}[t]
\centering
\IfFileExists{spectrum.pdf}{\includegraphics[width=\textwidth]{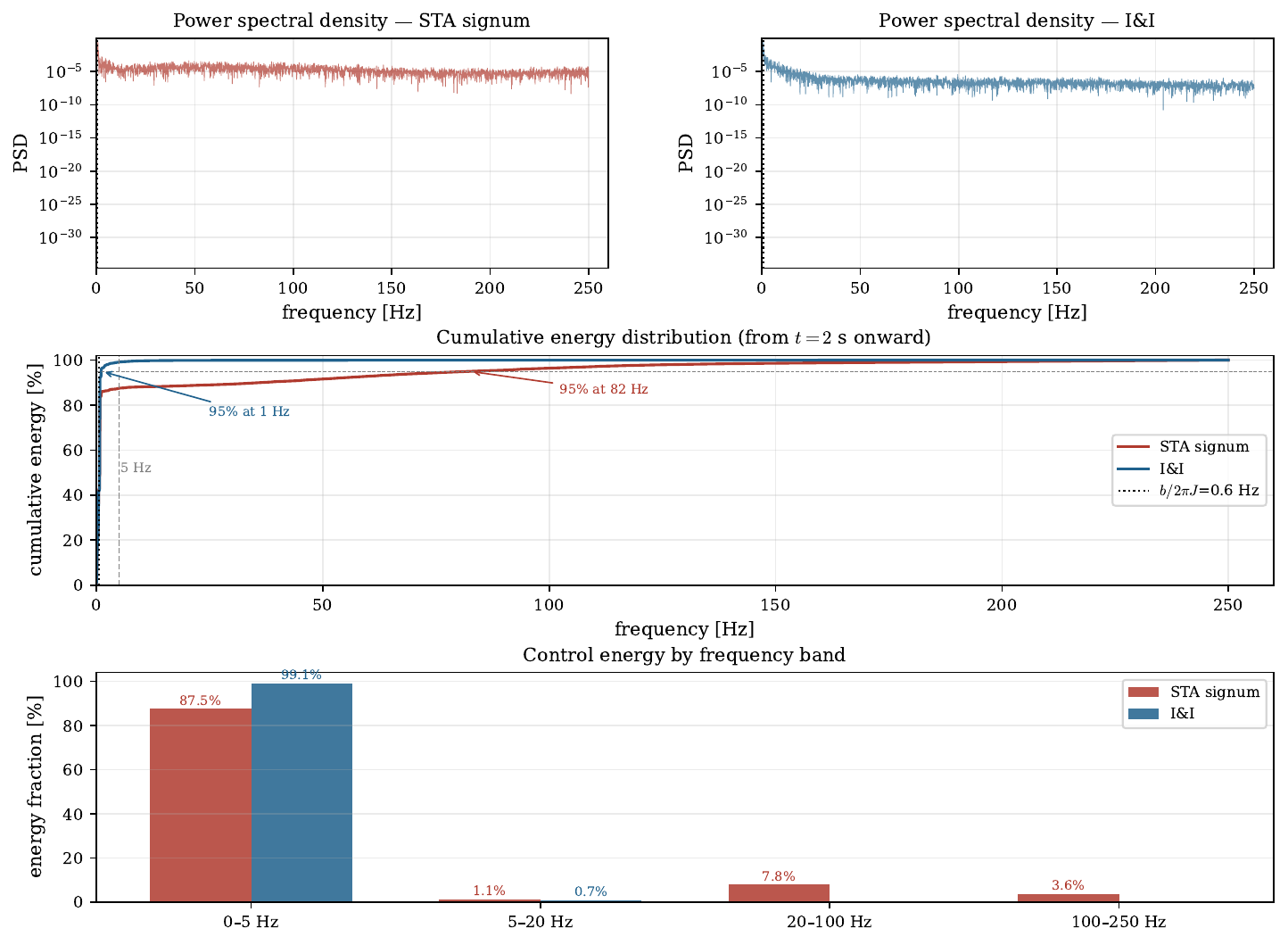}}{%
  \IfFileExists{spectrum.png}{\includegraphics[width=\textwidth]{spectrum.png}}{%
  \fbox{\parbox[c][60mm][c]{0.8\textwidth}{\centering\textbf{spectrum.pdf not found}}}}}
\caption{Fourier energy distribution of the control signal (from $t = 2$\,s onward).
The I\&I control concentrates
$99.1\%$ of its energy below $5$\,Hz. The STA signum control places $87.5\%$ below
$5$\,Hz, with $7.8\%$ in the $20$--$100$\,Hz range, and $3.6\%$ near the Nyquist frequency. The $11.4\%$ above
$5$\,Hz is the switching content visible in Fig.~\ref{fig:hw_compare}. The STA higher-frequency components sit
entirely above the mechanical bandwidth and are filtered by the motor winding
inductance and the $131\!:\!1$ gearbox before reaching the output shaft.
A \$10 H-bridge (L293) provides this filtering by construction.}
\label{fig:spectrum}
\end{figure}

Figure~\ref{fig:spectrum} quantifies what Figure~\ref{fig:hw_compare} shows
qualitatively. The STA control signal appears irregular because it switches at $500$\,Hz;
the spectrum reveals that $87.5\%$ of its energy is concentrated below $5$\,Hz,
with a structure identical to that of the I\&I control. 
The remaining $12.5\%$ is distributed across the $5$ to $250$\,Hz band,
with $8.9\%$ of the total energy corresponding to frequency modes of unstructured disturbances
(such as mechanical backlash and vibrations), located at around 131 times the mechanical frequency of 
the output shaft dynamics, at $b/(2\pi J) = 0.61$\,Hz.
The motor shaft cannot track this profile, as the electrical winding is unable to sustain it and the gearbox averages it out.
The difference in ISC between the two controllers ($0.74$ versus $0.77$) reflects this fact:
switching energy exists in the electrical domain, but it barely reaches the mechanical output.

This provides a quantitative answer to the concern regarding chattering raised in \cite{cervantes}. 
With gains exceeding the Levant-Moreno prescription by more than a factor of 100, 
the switching component would dominate the useful signal, and the motor would overheat, vibrate, and exhibit audible chattering.
At the appropriate level ($L = 40$), chattering accounts for approximately 3.6\% of the electrical energy and 0\% of the mechanical energy; 
of the total amount, 8.9\% is consumed in suppressing high-frequency components within unmodeled mechanical effects.
The distinction between chattering and switching at a frequency the plant cannot track is the distinction between excessive gains and correctly tuned gains.

\begin{rem}
    It is worth noting that the I\&I scheme makes more efficient use of control energy,
    since 99\% of it is used directly to actuate the output shaft mechanical subsystem.
    There is no energy in the unstructured disturbance band, given that the I\&I scheme simply ignores its existence
    and takes no measures to compensate for it.
    In a sense, this represents a clear advantage that was overshadowed by a rigged comparison
    rather than a scientifically valid one.
\end{rem}

\paragraph{A final observation on the nature of the actuator.}
Every result in \cite{cervantes}, and every result in this note, commands a DC motor
through a power driver. Whether that driver is a linear amplifier or a switching stage,
the voltage applied to the winding is a modulated signal: 
PWM, sigma-delta, or a current loop whose inner
controller is itself a switching law (a sliding mode scheme, again!).
The ``smooth'' I\&I control signal does not
reach the motor as a smooth voltage; it is discretized into switching pulses at the
driver's modulation frequency, typically $20$--$50$\,kHz. The signum-based STA
control adds its own switching content at $500$\,Hz, well below the driver's rate, and
as Fig.~\ref{fig:spectrum} shows, that content carries $3.6\%$ of the signal energy
and sits entirely above the mechanical bandwidth.
Therefore: condemning a $500$\,Hz switching
component in the control law, while accepting a $20$\,kHz switching component in the
driver, is not a coherent position. There is no smooth path from algorithm to torque
in a DC motor; there is only a choice of where the modulation occurs. The STA
observer places it where the physics accepts and requires it.

\subsection{The reported margin is an artifact of gain selection}

Table~\ref{tab:hw_kv} sweeps $k_v$ at fixed $k_p = 4000$, varying $\zeta$ from
$1.58$ to $0.08$. With signum at the perturbation bound, STA leads ISE at
$k_v = 200$, $60$ and $25$; \II{} leads only at $k_v = 10$, where both schemes
chatter due to the underdamped loop ($\zeta = 0.08$), not due to the switching
function. The $k_v = 25$ row is the cleanest single point: STA wins ISE by $11\%$ and
ISC by $6\%$, with no audible chattering.

\begin{table}[t]
\centering
\caption{Hardware, $k_p = 4000$ fixed, $k_v$ swept. STA uses signum with $L = 40$,
$\bar\theta = [b,c]^\top$.}
\label{tab:hw_kv}
\small
\begin{tabular}{@{}rrr rr r rr@{}}
\toprule
$k_v$ & $\zeta$ & $|p_1|$
  & STA $\ISE$ & \II{} $\ISE$ & ratio
  & STA $\ISC$ & \II{} $\ISC$ \\
\midrule
200 & 1.58 & 22.5
    & \textbf{0.00041} & 0.00070 & 0.59
    & 7.173 & \textbf{0.750} \\
 60 & 0.47 & 30.0
    & \textbf{0.00025} & 0.00025 & 1.00
    & 0.904 & \textbf{0.830} \\
 25 & 0.20 & 12.5
    & \textbf{0.00024} & 0.00027 & 0.89
    & \textbf{0.847} & 0.897 \\
 10 & 0.08 &  5.0
    & 0.00090 & \textbf{0.00045} & 2.00
    & 5.310 & \textbf{2.097} \\
\bottomrule
\end{tabular}
\end{table}

\section{Concluding remarks}

We do not dispute that the \II{} observer constitutes a robust design for mechanical systems with structured friction. In a plant whose dynamics conform to the regressor in Proposition 1 of \cite{romero25}, real-time adaptation absorbs model discrepancies that a fixed prior cannot track, and it does so smoothly. For a designer who has identified the plant, validated the friction model, and confirmed the regressor structure, \II{} represents a rational choice.

The super-twisting observer offers something different: it alleviates the modeling
effort. It requires no regressor structure, no friction parameterization, and no
identification campaign. It converges in finite time on any system whose perturbation
derivative is bounded, regardless of whether that perturbation is friction, backlash,
cogging, or something the designer has not named. Once converged, it practically delivers the same
steady-state tracking as \II{} (Panel~C of
Table~\ref{tab:results}).

What we dispute is the conclusion drawn in \cite{cervantes}. The comparison reported
there uses a pole placement ($\zeta = 13.7$) that makes the tracking-error dynamics
six times slower than the \II{} observer, ensuring the observer transient is
invisible; super-twisting gains that exceed the Levant--Moreno prescription by
$22$ to $105\times$, guaranteeing chattering; no performance index of any kind,
relying on visual inspection; and no boundary layer, omitting a standard technique.
Under these conditions any sliding-mode design will appear inadmissible, and any
model-based adaptive design will appear superior. The result characterizes the
experimental setup, not the methods.

Categorical conclusions of the form ``method $X$ is practically inadmissible''
propagate far beyond the readership that can evaluate them. A graduate student
selecting an observer, or an engineer choosing a technique for a product, will read
the abstract and act on it. When the supporting evidence does not survive a change of
pole placement, such a claim does active harm. An honest result showing a marginal or
zero advantage under controlled conditions is more useful to the community than a
dramatic one that does not replicate.

{\it 
While the physical demonstrations associated with \cite{cervantes} remain closed to community feedback, our experimental replication is made entirely transparent. We provide unedited, open-comment video proof demonstrating smooth tracking performance without audible chattering.
}

\bibliographystyle{unsrt}

\end{document}